  \documentclass[preprint]{aastex}

 \usepackage{longtable}
 
\newcommand{\Htwo}{H$_2$}            
\newcommand {\HI}     {\ion{H}{1}}      
\newcommand{\Lya}{Ly$\alpha$}

\newcommand{\etal} {et~al.}
\newcommand{\kms} {km~s$^{-1}$} 

\begin{document}

\title{Interstellar Bow Shocks around Fast  Stars \\
Passing through the Local Interstellar Medium} 

\author{J. Michael Shull}   
\affil{Department of Astrophysical and Planetary Sciences   \\
University of Colorado, Boulder, CO 80309 }  

\author{S. R. Kulkarni}  
\affil{Owens Valley Radio Observatory, 249-17 \\
California Institute of Technology, Pasadena, CA 91125} 

\email{michael.shull@colorado.edu, srk@astro.caltech.edu}


\begin{abstract}

 Bow-shocks are produced in the local interstellar medium by the passage of fast stars 
 from the Galactic thin-disk and thick-disk populations with velocities $V_* = $ 40--80~\kms.  
 Stellar transits of local \HI\ clouds occur every 3500--7000 yr on average and last
 between $10^4$ and $10^5$~yr.  There could be 10--20 active bow shocks around 
 low-mass stars inside clouds within 15 pc of the Sun. At local cloud distances of 3--10 pc, 
 their turbulent wakes have transverse radial extents $R_{\rm wake} \approx$ 100--300~AU ,
 angular sizes 10--100 arcsec, and \Lya \ surface brightnesses of  2-8 Rayleighs in gas 
 with total hydrogen density $n_{\rm H} \approx 0.1~{\rm cm}^{-3}$ and $V_* =$ 40-80~\kms.   
 These transit wakes may cover an area fraction 
 $f_A \approx (R_{\rm wake}/R_{\rm cl}) \approx 10^{-3}$ of local 
 \HI\ clouds and be detectable in IR (dust), UV (\Lya, two-photon), or non-thermal radio emission.   
 Turbulent heating in these wakes could produce the observed elevated rotational populations 
 of H$_2$ ($J \geq 2$) and influence the endothermic formation of CH$^+$ in diffuse interstellar 
 gas at $T > 10^3$~K.

\end{abstract} 


\section{Introduction}
 
 Recent observations have advanced our knowledge of gas clouds in the interstellar 
 medium (ISM) and the velocity dispersions of stars in the thin-disk and thick-disk populations.
 This is an opportune moment for a new look at the frequency and physical effects of bow 
 shocks produced by stars passing through low-density \HI\ clouds near the Sun. 
 Most of these transits will be by low-mass stars, although more distant examples of
  bow shocks around runaway O-type stars have been found in infrared emission 
 (Van Buren \& McCray 1988;  Peri \etal\ 2015; Kobulnicki \etal\ 2016).  The {\it Herschel} 
 far-infrared survey (Cox \etal\ 2012) identified numerous bow shocks around AGB stars 
 and red supergiants, including Betelgeuse (Decin \etal\ 2012).  Theoretical studies 
 of intermediate-velocity shocks (Kulkarni \& Shull 2023) suggest that the far-UV background 
 radiation may include contributions from \Lya\ and two-photon continuum emission from the 
 metastable (2s) state of \HI.  To excite the {\it 2s} and {\it 2p}  levels (10.2 eV) requires
 shock velocities $V_s  > 40$~km~s$^{-1}$ with post-shock temperatures 
 $T_s \approx (3 \mu V_s^2/16 k) \approx (36,300~{\rm K}) V_{40}^2$, where 
 $V_{40} = V_s / 40~{\rm km~s}^{-1}$.  Magnetic fields will reduce this post-shock
 temperature.

 In addition to their astronomical importance, fast stars passing through local clouds contribute
 to the thermal energy budget of interstellar gas in their transit wakes and extended regions.  
 Observations of interstellar H$_2$ in rotationally excited states with angular momentum quantum 
 number $J \geq 2$ suggest additional excitation mechanisms besides UV-pumping 
 (Gry \etal\ 2002; Nehm\'e \etal\ 2008; Ingalls \etal\ 2011; Shull, Danforth, \& Anderson 2021).   
 Elevated rotational populations could be produced by collisional excitation in warm molecular gas  
 ($T > 500$~K) heated by MHD turbulent dissipation (Moseley \etal\ 2021).  The observed 
 correlation of CH$^+$ with rotationally-excited \Htwo\ (Jensen \etal\ 2010) has led to similar 
 production schemes for CH$^+$ involving warm \Htwo\ (Falgarone \& Puget 1995; 
 Myers \etal\ 2015).  Local clouds of diffuse interstellar gas exhibit significant turbulent line widths 
 of 2--3 \kms\  (Linsky \etal\ 2022) derived by comparing the widths of absorption lines from atomic 
 species of different masses.  They found that the LISM is inhomogeneous on length scales
 less than 5100~AU.  The inferred temperature distribution appears Gaussian, with some sight
 lines reaching $T \approx 12,000$~K. Table~1 lists a number of local ``fast stars", together 
 with their velocities and properties of the LISM (gas temperatures and turbulent velocities).
 
Strong shocks will be produced by stars moving at velocities (40--70~\kms) through the
warm neutral medium of local clouds, in excess of the sound speed $c_s$ (or magnetosonic 
wave speed $v_m$).  For magnetized gas, magnetosonic waves are characterized by a 
combination of adiabatic sound speed,  
$c_s = (5 kT/3 \mu)^{1/2} = (9.30$~km~s$^{-1})T_{8000}^{1/2}$, and Alfv\'en velocity, 
$v_A =(5.53~{\rm km~s}^{-1}) (B/3\,{\mu G}) n_{\rm H}^{-1/2}$.
Here, we scaled to gas temperature $T \approx (8000~{\rm K})T_{8000}$, total hydrogen density 
$n_{\rm H}$ (in cm$^{-3}$), mean particle mass $\mu = 1.27 m_H$ for $y \equiv$ He/H = 0.1, and 
magnetic field strength $B$.   Absorption-line studies (Frisch \etal\ 2011; Linsky \etal\ 2022) of 
clouds in the local interstellar medium (LISM) suggest $n_{\rm H} \approx 0.1$~cm$^{-3}$ and 
$B \approx$ 3--4~$\mu$G (McComas \etal\ 2012; Zank \etal\ 2013).  For wave propagation 
perpendicular to the magnetic field, the fast-mode magnetosonic wave speed, 
$v_m = (v_A^2 + c_s^2)^{1/2}$.   Considering spatial variations in LISM parameters
 ($n_H$, $T$, $B$), magnetosonic wave speeds in the LISM could range from 20 to 30~\kms, 
 with further uncertainties arising from propagation angle and
 ionization fraction\footnote{Ionization fractions are important at the shock interface, owing 
 to charge exchange reactions with neutrals.  Scherer \& Fichtner (2014) also pointed out the 
 importance of including both He$^+$ and H$^+$ in determining the speed of Alfv\'en waves.  
 For wave propagation (vector ${\bf k}$ at angle $\phi$ to the magnetic field, the fast-mode
 (Kulsrud 2005) dispersion relation 
$(\omega/k)^2 = (v_A^2 + c_s^2)/2 + \frac{1}{2} [(v_A^2 - c_s^2)^2 + 4 v_A^2 c_s^2 \sin^2 \phi]^{1/2}$
defines the phase and group velocities of the wave packets.   Near the front of the bow shock,
we expect that $\phi \approx 90^{\circ}$ (${\bf k} \perp {\bf B}$).  Away from the forward region, 
the fast-mode wave speed could be somewhat smaller than $(v_A^2 + c_s^2)^{1/2}$.}.

 Observations of the kinematics of the Milky Way disk suggest that populations of older stars 
 in the Galactic disk have supersonic and super-Alfv\'enic velocities relative to the LISM.  
 The large stellar velocity dispersions measured in the Galactic thin disk (Nordstrom \etal\ 2004; 
 Anguiano \etal\ 2020) are likely the result of dynamical heating processes that increase the
 stellar velocities. Massive Galactic objects (spiral arms, giant molecular clouds, bars,
 infalling dwarf satellites) perturb the motions of the newborn stars, causing their random 
 velocities to increase with time.  A recent study (Viera \etal\ 2022) using {\it Gaia}-DR3 proper 
 motions found two kinematically distinct Gaussian components: a thin disk with velocity 
 dispersions ($\sigma_R$, $\sigma_{\phi}$, $\sigma_Z$) = 30, 21, and 11 \kms\ and a thick 
 disk with dispersions of  49, 35, and 22 \kms.  These correspond to 2D (in-plane) dispersions of 
$\sigma_{\rm 2D} = 37$~\kms\ (thin disk) and 60~\kms\ (thick disk).  The vertical (out-of-plane) 
dispersions (11 and 22~\kms) and total 3D dispersions (38 to 64~\kms) also differ between the 
components.  These velocities are considerably larger than those of stars forming inside dense 
molecular gas,  seen in stellar associations and star clusters with radial velocity dispersions 
$\sigma_{\rm 1D} = $1--3 km~s$^{-1}$.  Proper motion measurements with {\it Gaia} 
(Kuhn \etal\ 2019; Shull, Darling, \& Danforth 2021) show that young star clusters are expanding 
at velocities $\sim 0.5$~km~s$^{-1}$ and will disperse in several Myr after the gas is expelled.

\section{Stellar Bow Shocks in the Local ISM}  

\subsection{Star-Cloud Encounters} 

Because of the large velocity dispersions for stars in Galactic disk populations, \HI\ clouds in the 
LISM will have frequent encounters with stars at velocities 40--70 km~s$^{-1}$.  Absorption-line 
studies along 157 stellar sight lines (Redfield \& Linsky 2008; Linsky \etal\ 2019) identified numerous 
warm \HI\ clouds within 15 pc of the Sun. The nearest structures within $d \sim 4$~pc are labeled 
the Local Interstellar Cloud (LIC), the G~Cloud, and the Blue Cloud.  Redfield \& Linsky (2008) 
 estimated that between 6 and 19\% of the LISM volume is filled with warm \HI\ gas.  An updated 
 analysis (Linsky \etal\ 2022) found that clouds within 4~pc are more tightly packed, with mean 
 hydrogen density $\langle n_{\rm HI} \rangle \approx 0.1$~cm$^{-3}$ and typical volume factors 
 $f_V \approx 0.38-62$.  For local clouds beyond 4~pc, they find $f_V < 0.4$. 
 
We now estimate the rate at which passing stars encounter warm \HI\  gas clouds in the LISM 
from cloud cross sections, stellar space densities, and velocity dispersions. If the region
within distance $R_{\rm LISM}$ of the Sun is filled by $N_{\rm cl}$ spherical clouds of 
radius $R_{\rm cl}$, the volume filling factor is $f_V = N_{\rm cl} (R_{\rm cl} / R_{\rm LISM})^3$,
and  the area covering factor is $f_A =  N_{\rm cl} (R_{\rm cl} / R_{\rm LISM})^2$.  These factors 
are related by $f_A = f_V (R_{\rm LISM}/R_{\rm cl})$ for spheres.   More generally, $f_A$ and 
$f_V$  depend on cloud geometry, although the non-spherical or filamentary shapes of some
local clouds do not change these estimates substantially.  For 15 clouds within 10 pc, the above
approximation suggests mean cloud radii, 
$R_{\rm cl} = (10~{\rm pc})(f_V / N_{\rm cl})^{1/3} \approx$ 2--3~pc and area covering factors
$f_A \approx$ 0.7--1.0 for $f_V =$ 0.15--0.30.  

The star-cloud encounter rate depends on the local density of stars, both those currently 
seen within the local clouds and the stars in the disk populations that transit the LISM at 
high velocity.  The recent Gaia-DR3 survey of the local volume within 10~pc (Reyl\'e \etal\ 2022) 
found 541 stars in 336 systems, of which 286 were luminous stars (we do not include the 85 
brown dwarfs in the survey).  This corresponds to a mean density of $0.065~{\rm pc}^{-3}$, 
similar to that quoted in Table 4.5 of Mihalas \& Binney (1981).
The normalized densities of stars in the thin-disk and thick-disk populations should sum to
similar values, although there is some disagreement as to their relative local densities 
(Anguiano \etal\ 2020;  Everall \etal\ 2022; Viera \etal\ 2023).  Not all of these stars will have
velocities sufficient to produce the type of shocks discussed here.  Only a fraction will
have the high velocities needed to exceed magnetosonic speeds.  The radial extent of the
astrospheres, bow shocks, and turbulent  wakes depend on the presence of a moderately 
strong stellar wind (Herbst \etal\ 2020).  

To be conservative, we scale our encounter rate to a local density $n_* = 0.01~{\rm pc}^{-3}$ 
of stars of mean velocity $V_* = (40~{\rm ~km~s}^{-1}) V_{40}$, passing through an extended 
LISM  with $R_{\rm LISM} = (15~{\rm pc})R_{15}$.  The collision rate with local clouds is
\begin{equation} 
     k_{\rm coll} = n_* (\pi R_{\rm LISM}^2 ) f_A V_* \approx 
     (3 \times 10^{-4}~{\rm yr}^{-1}) \, V_{40} R_{15}^{2} f_A \; . 
\end{equation}
For $f_A \approx$ 0.5--1.0, the encounter rate is one collision every 3500--7000 yr.  
A typical stellar crossing time through a cloud of mean length 
$\ell  \approx 4R_{\rm cl} / 3 \approx 3$~pc is
\begin{equation}
   t_{\rm cross} \approx  \ell  / V_*  \approx (70,000~{\rm yr}) (\ell /3~{\rm pc})  V_{40}^{-1}   \;  ,
\end{equation}
and the number of active bow-shock encounters within 15 pc is
\begin{equation}
   N_{\rm bow} = k_{\rm coll} \, t_{\rm cross} \approx 20 \, (\ell /3~{\rm pc}) f_A R_{15}^2   \; .   
\end{equation}
The chances of detection could be significantly higher for larger cloud complexes beyond 
the LISM (Frisch \etal\ 2011).  However, the local clouds lie inside a low-density cavity, and
the nearest star-forming clouds are more distant, including gas in the Taurus molecular
cloud (140~pc) and the Scorpius-Centaurus association (130~pc).  

\subsection{Estimated Bow Shock Sizes}  

The two {\it Voyager} missions passed through the solar-wind termination shock at distances 
of 94 and 84 AU (Stone \etal\ 2005;  Richardson \etal\ 2008; Stone \etal\ 2008).  Low-mass stars 
like the Sun, with weak winds and low mass-loss rates, are expected to produce astrospheres
preceded by bow shocks with transverse radial extents $R_{\rm wake} \approx$100--300~AU, 
similar to that of the Sun's heliosheath and termination shock\footnote{Observations from the 
{\it Interstellar Boundary Explorer} (McComas \etal\ 2012; Zank \etal\ 2013; McComas \etal\ 2019) 
suggest that the solar wind termination may be in a ``bow wave" rather than a strong shock, 
owing to the influence of inclined magnetic fields in the heliosphere and LISM.  
This interpretation has been questioned by Scherer \& Fichtner (2014), who noted the 
importance of including both He$^+$ and H$^+$ in the speeds of Alfv\'en and magnetosonic 
waves.} of the solar wind.  These estimates are consistent with results from an HST survey 
(Wood \etal\ 2021) of astrospheric Ly$\alpha$ absorption toward nearby M-dwarf stars.  
For stars with super-magnetosonic motions through the ISM, the hydrogen wall and 
bow shock in MHD models (e.g., Zank \etal\ 2013) precede the wind termination shock
by a substantial factor that depends on the stellar speed and magnetosonic Mach number 
relative to the surrounding medium. This warm, extended gas is influenced in density,
temperature,  and wave speeds.  The trailing wakes typically have transverse radial extent 
2--3 times the size of the forward radius of the termination shock.
Most low-mass stars have mass-loss rates comparable to (and sometimes less than) that of 
the Sun, $\dot{M} \approx 4 \times 10^{-14}~M_{\odot}~{\rm yr}^{-1}$.   At the typical 3--10~pc 
distances to the local interstellar clouds, the bow-shock angular sizes could range from 
10--100~arcsec.  More distant bow shocks would be smaller (0.1 arcsec at 1~kpc) and probably 
undetectable.  

A few, more distant fast-moving OB stars with much stronger winds have observed arc-like 
features, attributed to bow shocks of parsec size (Van Buren \& McCray 1988; 
Mackey \etal\ 2016).   An O-star moving at velocity $V_* \approx 40~{\rm km~s}^{-1}$ would 
produce a wind-termination shock of approximate size,  
 \begin{equation}  
   r_b \approx \left[\frac {\dot{M}_w V_w}{4 \pi \mu n_{\rm H}} \right]^{1/2}  V_*^{-1} 
       \approx  (1.0~{\rm pc}) \, \dot{M}_{-6}^{1/2} n_{\rm H}^{-1/2} 
        \left( \frac {V_w}{1000~{\rm km~s}^{-1}} \right)^{1/2}  
        \left( \frac {V_*}{40~{\rm km~s}^{-1}} \right)^{-1}   \; .
 \label{eq:BowRadius}
\end{equation} 
This termination shock in O-star winds will be preceded a wind-driven bubble of
high-pressure gas (Weaver \etal\ 1977; Mackey \etal\ 2016), whose shape can be distorted 
by supersonic stellar motion.  Here, we express the stellar mass-loss rate as 
$\dot{M} = (10^{-6}~M_{\odot}~{\rm yr}^{-1}) \dot{M}_{-6}$ and equate the wind ram pressure, 
$\dot{M}_w V_w / 4 \pi r^2$, with the pressure, $\rho_{\rm ISM} V_*^2$, of the ISM of mass 
density $\rho_{\rm ISM} = 1.4 m_{\rm H} n_{\rm H}$.  At $d \approx 1$~kpc, the angular sizes
would be $\theta_b = r_b/d \approx 10^{-3}$ rad  $(3.4'~d_{\rm kpc}^{-1})$.  The above estimate 
of $r_b$ is uncertain by factors of 1.5--2.0 when one includes the effects of magnetic fields.  

\subsection{Estimated Luminosity and Surface Brightness} 

Intermediate velocity shocks moving into \HI\ gas will produce optical and UV emission lines that could be
signatures of fast-star passages through local clouds.  These include \Lya, metal-ion emission lines,
and two-photon continuum from the metastable (2s) state of \HI\ (Shull \& McKee 1979).  An analysis of the 
bow-shock luminosity and surface brightness requires calculations of the photon flux, $\Phi_0$, which is 
proportional to the flux, $n_{\rm H} v_s$, of hydrogen atoms entering the shock front.  The radiative shock 
models (Shull \& Silk 1979) of primeval material (H and He) found photon fluxes that can be fitted
 to expressions rising steeply for $V_s$ between 15 and 100~km~s$^{-1}$, 
\begin{eqnarray}
    \Phi_0(2\gamma)        &=& (2.2 \times 10^4~{\rm phot~cm}^{-2}~{\rm s}^{-1} )
                      (n_{\rm H} / 0.1~{\rm cm}^{-3}) V_{40}^{2.18}   \; , \\
    \Phi_0({\rm Ly}\alpha) &=& (2.7 \times 10^5~{\rm phot~cm}^{-2}~{\rm s}^{-1} )
                      (n_{\rm H}/0.1~{\rm cm}^{-3}) V_{40}^{2.12}   \; . 
\end{eqnarray}
We use these expressions to estimate the luminosities of \Lya\ and two-photon emission produced in a 
bow shock of radius $r_b$ from a low-mass star located at distance $d$.   Because of the approximate
nature of these estimates, we consider plane-parallel shocks at small angles from the forward direction
of the star.  More careful 3D models with MHD codes would clearly be warranted in the future.  Most of 
the emission comes from a polar cap with opening angle 
$\theta_{\rm cap} \approx 30^{\circ}$ relative to the stellar velocity vector.  This cap occupies a fraction 
$(1 - \cos \theta_{\rm cap})$ of the $2 \pi r_b^2$ area of the forward hemisphere, with a photon luminosity,
\begin{equation}
   S_{\rm phot} = 2 \pi  r_b^2  \Phi_0 (1 - \cos \theta_{\rm cap})  \;  ,
\end{equation}
and its projected solid angle on the sky is approximately 
\begin{equation}
  \Omega_{\rm arc}  \approx \frac {2  r_b \theta_{\rm cap} (\Delta r)}{ d^2} = 
        \left( \frac {2 r_b^2}{d^2} \right) \theta_{\rm cap} (1 - \cos \theta_{\rm cap})  \; .
\end{equation} 
Here, we adopt an angular length, $2 \theta_{\rm arc} =  2 \theta_{\rm cap}$, for the arc
(angles in radians) and a projected arc width $\Delta r =  r_b (1 - \cos \theta_{\rm cap})$. 
Using eq. (7), the arc surface brightness is
\begin{equation}
   B_{\rm arc} = \frac {S_{\rm phot}/4 \pi d^2} {\Omega_{\rm arc} } 
                      = \frac {\Phi_0} {4 \theta_{\rm cap} } \; .
\end{equation}
As expected for optically thin emission, the brightness is independent of both $d$ and $r_b$.  

For solar-type stars passing through the LISM  at $V_* \approx (40$~km~s$^{-1})V_{40}$, we adopt 
$r_b = 100$~AU, $n_{\rm H} \approx 0.1$~cm$^{-3}$, and  
$\theta_{\rm cap} \approx 30^{\circ}$ to find 
\begin{eqnarray} 
     S_{\rm phot} (2\gamma)        &=& (4.2 \times 10^{34}~{\rm phot~s}^{-1}) 
          \left( \frac {n_{\rm H}}{0.1~{\rm cm}^{-3}} \right)  V_{40}^{2.18}   \; ,  \\
     S_{\rm phot}({\rm Ly}\alpha)  &=& (5.1 \times 10^{35}~{\rm phot~s}^{-1}) 
          \left( \frac {n_{\rm H}}{0.1~{\rm cm}^{-3}} \right)  V_{40}^{2.12}  \; . 
\end{eqnarray} 
With the 10.2 eV energies of the upper ({\it 2s}, {\it 2p}) levels, these luminosity coefficients sum 
to $9 \times 10^{24}$ erg~s$^{-1}$, comparable to the kinetic energy of interstellar gas entering
the bow shock over the cap portion of the front,
\begin{equation}
   \frac {\rho_{\rm ISM} V_s^3}{2} (2 \pi r_b^2) (1 - \cos \theta_{\rm cap}) \approx
        (1.4 \times 10^{25}~{\rm erg~s}^{-1})  \left( \frac {n_{\rm H}}{0.1~{\rm cm}^{-3}} \right) V_{40}^3  \; . 
\end{equation} 
Additional heating of the dust in the gas assembled by the bow front will come from stellar radiation,
particularly in the case of OB stars.  The \Lya\ photons will scatter in the  low-density local \HI, but
owing to low \HI\ column densities only a small fraction will be absorbed by dust grains.  However,
some of these bow shocks may also be seen in infrared emission, as observed for OB-stars and
young-star outflows with the {\it Spitzer Space Telescope} (Winston \etal\ 2012; Kobulnicki \etal\ 2016) 
and the {\it Wide-field Infrared Survey Explorer} (Peri \etal\ 2015).  

Most of the surface brightness of the bow shock comes from the material directly in front of 
the star, visible on the sky as a small arc of emission (Kobulnicky \etal\ 2016).
Adopting  $\theta_{\rm cap} = 30^{\circ}$  ($\pi/6$ radians) and $\Delta r / r_b \approx 0.13$, 
and using  equations (5) and (6) for $\Phi_0$, we estimate photon surface brightnesses,
$B_{\rm arc} \approx (3 \Phi_0/2 \pi)$ of the arc,
\begin{eqnarray}
    B_{\rm cap}(2\gamma)       & \approx &  (10,500~{\rm phot~cm}^{-2}~{\rm s}^{-1}~{\rm sr}^{-1})
              V_{40}^{2.18} \left( \frac {n_{\rm H}}{0.1~{\rm cm}^{-3} } \right)   \; ,  \\
    B_{\rm cap}({\rm Ly}\alpha) & \approx & (129,000~{\rm phot~cm}^{-2}~{\rm s}^{-1}~{\rm sr}^{-1})
              V_{40}^{2.25} \left( \frac {n_{\rm H}}{0.1~{\rm cm}^{-3} } \right)  \;  .
\end{eqnarray}  
The above brightness coefficients (in units photons cm$^{-2}$ s$^{-1}$ sr$^{-1}$) can also be 
expressed as 0.13 and 1.6 Rayleighs.  Thus, for stars in the thin-disk and thick-disk populations,
with rms velocities of 37~\kms\ and 60~\kms\ respectively, the \Lya\ surface brightnesses could 
range from 2--4 Rayleighs in gas with $n_{\rm H} \approx 0.1$~cm$^{-3}$.  
Even greater brightnesses might appear around several of the fast local stars 
listed in Table 1.

\section{Turbulent Heating of the LISM gas} 

The LISM clouds penetrated by transiting low-mass stars will be affected by turbulence 
generated by transiting fast stars at various shocks and in the wake.
In clouds of pathlength $\ell =$ 2--3~pc, stars moving at 40--80~\kms\
will spend 30,000--100,000 years, leaving wakes of dynamically disturbed gas.   Owing to 
downstream flows from the bow shock, the transverse sizes of the wakes will be larger than 
the forward termination shock and astrosheath (e.g., Mac~Low \etal\ 1991; Mackey \etal\ 2016).
The fractional volume of the cloud affected  by the turbulent wake is
\begin{equation}
   f_V  \approx \left( \frac {R_{\rm wake}} {R_{\rm cl}} \right)^2  \approx (2 \times 10^{-6})
         \left[ \frac {R_{\rm wake}/300~{\rm AU}} {R_{\rm cl}/{\rm pc}} \right]^2   \; .
\end{equation}
Note that $R_{\rm wake}$ is the transverse size of the post-bowshock flow, not its length.
The total affected portion of local clouds  includes both the direct ``bore holes" and 
extended regions influenced by heat generated by MHD wave damping.    As described by 
Spitzer (1982) and Ferri\'ere \etal\ (1988), the waves arise from reflected shocks produced by 
interactions with density irregularities encountered by the bow shock.  After a nonlinear cascade 
to smaller eddies, much of the turbulence will decay into heat, on an eddy-turnover time,
\begin{equation}
     t_{\rm eddy}\approx R_{\rm eddy} / V_{\rm eddy} \approx (10^3~{\rm yr}) 
     \left( \frac  {r_b}{100~{\rm AU}} \right)  \left( \frac {3~{\rm km/s} }  {V_{\rm eddy} } \right)  \; . 
\end{equation}
Here, we assumed the initial eddy to have a characteristic size 
$R_{\rm eddy} \approx  r_b = 100~{\rm AU}$ at the termination shock scale and an initial turbulent 
velocity $V_{\rm eddy} \approx 3~{\rm km~s}^{-1}$.  After cascading to smaller scales through
nonlinear effects, the turbulent energy is dissipated.  The turbulent heating rate can be estimated 
by dividing the kinetic energy density ($\rho_{\rm gas} V_{\rm eddy}^2$) by the eddy turnover time,
\begin{equation}
    \epsilon_{\rm turb} \approx \rho_{\rm gas} V_{\rm eddy}^3 / R_{\rm eddy}
    \approx (4\times10^{-23}~{\rm erg~s}^{-1}) n_{\rm H}  \left( \frac {r_b}{100~{\rm AU}} \right)  
     \left( \frac {3~{\rm km~s}^{-1}}  {V_{\rm eddy}} \right)^3  \;  , 
\end{equation}
where we assumed $\rho_{\rm gas} = 1.4 n_{\rm H} m_{\rm H}$.  The turbulent heating time would be
\begin{equation}
    t_{\rm heat} \approx \frac {3nkT/2} {\epsilon_{\rm turb}}
        \approx (10^3~{\rm yr}) \left( \frac {r_b}{100~{\rm AU}} \right)^{-1}
           \left( \frac {V_{\rm eddy}} {3~{\rm km~s}^{-1}} \right)^{-3} 
          \left( \frac {T}{6000~{\rm K}} \right) \; .
\end{equation}
Because this time is much shorter than the stellar crossing time through the cloud, the thermal state 
of gas in the bow-shock wake will be time dependent.  The situation will involve transient heating 
over multiple time scales, with $t_{\rm heat} <  t_{\rm cool}$ and $t_{\rm cool} \approx t_{\rm cross}$, 

The post-shock wake will cool gradually, as the turbulent eddies quickly cascade down to smaller scales.  
Gas temperatures will rise transiently to $T \approx 20,000-30,000$~K, where \HI\ can be collisionally 
ionized.  Radiative cooling by \HI\ (\Lya) and metal forbidden lines will reduce the temperature to 
$T \leq 10^4$~K on an isobaric cooling timescale, 
\begin{equation}
   t_{\rm cool} = \frac {5n_{\rm tot} kT/2} {n_e n_H {\cal L}(T) }
       \approx (50,000~{\rm yr}) \left( \frac {n_{\rm H}} {0.1~{\rm cm}^{-3}} \right)^{-1}  
       \left( \frac {T} {20,000~{\rm K}} \right) {\cal L}_{-22}^{-1}   \;  , 
\end{equation}
comparable to the stellar crossing time.  We assume a partially ionized medium with 
$n_e \approx 0.5n_{\rm H}$ and $n_{\rm tot} \approx 1.6 n_{\rm H}$, and a cooling rate coefficient
${\cal L} \approx (10^{-22}~{\rm erg~cm}^3~{\rm s}^{-1}) {\cal L}_{-22}$ at 20,000~K. Analyzing 
the time-dependent evolution of ionization and temperature behind bow shocks is a problem 
in hydrodynamics and thermodynamics worth further study.   
The effects of magnetic fields in the turbulent wakes (Baalmann \etal\ 2021 add to the 
complexities, including reconnection and dissipation of magnetic field energy.

In our model we have ignored cooling through all elements other than hydrogen and ignored external 
heat inputs.  By $10^4$\,K the cooling power from hydrogen decreases with temperature and will 
compete with cooling by forbidden and fine structure lines of metals.  As cooling proceeds, the internal 
heat is proportionally reduced and heating  by external agents (e.g., ionization by cosmic rays, 
photoelectric dust heating, ionizing soft X-ray radiation) will become important.  The recombination 
time scale is about $10^5n_{\rm H}^{-1}$~yr. The ionization time per atom from the diffuse EUV field 
which ionizes the warm ionized medium is $10^6$~yr. 

\section{Conclusion and Future Prospects}
 \label{sec:ConclusionProspects}

This paper analyzes the interstellar bow-shocks in the LISM generated around passing stars from 
the Galactic thin disk and thick disk with velocities of 40--70~\kms.  We estimate that bow shocks 
transit the local \HI\ clouds every 3500--7000 yr.  On average, there could be 
10--20 active bow shocks inside clouds within 15 pc of the Sun, with stellar transit times varying from 
$10^4$--$10^5$~yr.    At typical local cloud distances of 3--10 pc, the bow shocks and astrospheres 
around  low-mass stars likely have 100--300~AU transverse radial extent and angular sizes of 
10--100 arcsec.  Their turbulent transit wakes of passing stars might cover an area fraction 
$f_A \approx (R_{\rm wake} / R_{\rm cl})^2 \approx 10^{-3}$ of nearby clouds.  They could explain
the 2--3~\kms\ turbulent widths and temperature variations on scales less than 5100~AU in LISM 
clouds (Linsky \etal\ 2022), as well as elevated rotational populations of H$_2$ and endothermic 
production of CH$^+$.  Bow shocks around rarer massive stars and AGB stars would produce 
shocks of parsec scale and angular sizes  of several arcmin (at $\sim1$~kpc). 

Detecting the low-mass bow shocks will require wide-field surveys, either in the infrared (dust
emission) or in the ultraviolet (for \Lya\ and two-photon emission).  The IR detections could be a 
task for the {\it WISE} data base, and \Lya\ emission could be a project for the Alice spectrograph 
aboard the New Horizons (NH) spacecraft during its extended mission.  During six great-circle scans 
(at 40-47 AU) the NH spectrograph observed interplanetary \Lya\ (Gladstone \etal\ 2021) with a 
uniform background of $43\pm3$ Rayleighs over the sky, interpreted as the Galactic \Lya\ background.  
There is also resonantly scattered solar \Lya, which at the current NH distance (56 AU from the Sun) 
has fallen to about the same level.  Thus, the typical brightness in any direction (away from the Sun) 
would be 80--100 R.  Our calculations of far-UV radiation from the forward caps of 40--80 \kms\ bow 
shocks suggest that arc-like structures around low-mass stars transiting local clouds with 
$n_{\rm H} \approx 0.1~{\rm cm}^{-3}$ should have surface brightnesses of 2--8 Rayleigh in \Lya\ 
and 0.1--0.4 Rayleigh in the UV two-photon continuum.  These could possibly be seen with NH 
and with slit spectroscopy on the proposed UVEX mission (Kulkarni \etal\ 2021).  

Examples of bow shocks around runaway O-type stars and AGB stars have already been found in 
infrared emission (Kobulnicki \etal\ 2016; Peri \etal\ 2015; Cox \etal\ 2012).   Wide-angle searches 
in \Lya, far-UV continua, and infrared emission for lower luminosity examples could detect the more 
numerous cases of low-mass stars transiting the LISM. Their signatures might also be detectable in 
non-thermal radio emission from particles accelerated at the termination shock and bow-shock vicinity.   
Although the predicted stellar transit wakes have small covering fractions of LISM clouds, both by 
volume and on the sky, the effects of down-stream turbulent dissipation, MHD wave heating, 
and energetic particle acceleration could be more widespread throughout the local clouds.

\medskip 

\begin{acknowledgements}

We thank Jeffrey Linsky and Seth Redfield for discussions on gas structures in the local interstellar
medium, Randy Gladstone for insights on \Lya\ scanning observations with the New Horizons spacecraft, 
and Brian Wood for providing characteristics of the winds and astrospheric absorption around nearby 
solar type stars.  We also thank the anonymous referee for introducing us to important papers 
on MHD models of astrospheres.  

\end{acknowledgements} 



\newpage


\begin{deluxetable}{llccrcc}
\tablecolumns{7}
\tabletypesize{\scriptsize}
\tablenum{1}
\tablewidth{0pt}
\tablecaption{Nearby Fast Stars\tablenotemark{a} }

\tablehead{
  \colhead{Star}
&\colhead{SpT} 
&\colhead{$V_{\rm ISM}$}
&\colhead{$V_{\rm rad}$}
&\colhead{$d$}
&\colhead{$T_{\rm gas}$} 
&\colhead{$V_{\rm turb}$}  
\\
   \colhead{}
& \colhead{}     
& \colhead{(km/s)}
& \colhead{(km/s)}
& \colhead{(pc)}
& \colhead{(K)} 
& \colhead{(km/s)} 
}

\startdata
61~Cyg~A           &  K5~V       &  86  &  -65.2      & 3.50    &   6500   &   2.08      \\
GJ~887               & M2~V       &  85  &  +8.8        &  3.29   &  6000    &   1.65    \\
GJ~436               & M3~V       &  79  &  +9.6        &  9.78   &  \dots    &   \dots    \\
GJ~205               & M1.5~V    &  70  &  +8.6        &  5.70   &  4500    &   2.37     \\
$\epsilon$~Ind    & K5~V        &  68  &  -40.0       &  3.64   &  8340    &   1.97     \\
61~Vir                 & G6.5~V     & 51  &   -7.8         &  8.53   &  6600    &   1.60     \\
GJ~892               & K3~V        & 49  &  -18.5        &  6.54   &  \dots    &   \dots     \\
GJ~860A             & M3~V       & 47  &  -24.0        &  4.01    &  4780    &   2.24     \\
EV~Lac               & M3.5~V    & 45  &  +0.3         &  5.05    &  3030    &   2.21     \\
$\pi^1$~UMa       & G1.5~V    & 43  &  -12.8        & 14.4     &  2450    &   2.47     \\
36~Oph~A           & K2~V       & 40  &  +0.4         &  5.95    &  5870    &   2.33     \\
GJ~173                & M1~V      & 38  &  -6.8          & 11.2      &  \dots    &   \dots     \\
$\delta$~Eri         & K0~IV      & 37  &  -6.2          &  9.09     &  3650    &   2.25     \\
$\xi$~Boo~A        & G7~V       & 32  &  +1.6         &  6.71     &  5310    &  1.68     \\
GJ~338~A           & M0~V       & 29  & +11.5        &  6.33     &  6650    &  2.29     \\
$\delta$ Pav        & G8~IV      & 29  &  -21.5        &  6.10     &  9100    &  2.44     \\          
GJ~15~A             & M2~V       & 28  & +11.7        &  3.56     &  7800    &  2.81     \\ 
$\epsilon$~Eri     & K2~V        & 27  & +16.4        &  3.22     &  7410    &  2.03     \\
$\alpha$~Cen~B & K1~V       & 25  &  -22.6        &  1.35     &  5500   &  1.37     \\                  
Sun                      & G2~V       & 23 &   \dots        &  \dots    &  6400    &  \dots     \\
                            &                  &      &                   &             &              &                \\
{\bf  Means}         &                  &      &                   &             &  6638    &  2.54       \\      
\enddata

\tablenotetext{a} {~List of local ``fast stars" with spectral types and Gaia parallax distances.  
$V_{\rm ISM}$ is the ISM flow speed seen by the star in the stellar rest frame (Wood \etal\ 2021),
and $V_{\rm rad}$ is the star's heliocentric radial velocity.  The last two columns list gas 
temperatures and turbulent velocities, inferred from an ensemble of absorption-line 
widths of hydrogen (both H~I and D~I) and various metal ions (Linsky \etal\ 2022). } 

\end{deluxetable}


\end{document}